\documentclass[11pt]{article}
\usepackage{amssymb}
\newcommand{\equal}{\!\!\!&=&\!\!\!}
\newcommand{\newequiv}{\!\!\!&\equiv&\!\!\!}
\begin{document}
\abovedisplayshortskip 12pt
\belowdisplayshortskip 12pt
\abovedisplayskip 12pt
\belowdisplayskip 12pt
\baselineskip=18pt
\title{{\bf Deformed Harry Dym and Hunter-Zheng Equations}}
\author{J. C. Brunelli$^{a}$, Ashok Das$^{b}$ and Ziemowit Popowicz$^{c}$  \\
\\
$^{a}$ Departamento de F\'\i sica, CFM\\
Universidade Federal de Santa Catarina\\
Campus Universit\'{a}rio, Trindade, C.P. 476\\
CEP 88040-900\\
Florian\'{o}polis, SC, Brazil\\
\\
$^{b}$ Department of Physics and Astronomy\\
University of Rochester\\
Rochester, NY 14627-0171, USA\\
\\
$^{c}$ Institute of Theoretical Physics\\
University of Wroc\l aw\\
pl. M. Borna 9, 50-205, Wroc\l aw, Poland\\}
\date{}
\maketitle

\begin{center}
{ \bf Abstract}
\end{center}

We study the deformed Harry Dym and Hunter-Zheng equations with
two arbitrary deformation parameters. These reduce to various
other known models in appropriate limits. We show that both these
systems are bi-Hamiltonian with the same Hamiltonian structures.
They are integrable and belong to the same hierarchy corresponding
to positive and negative flows. We present the Lax pair
description for both the systems and construct the conserved
charges of negative order from the Lax operator. For the deformed
Harry Dym equation, we construct the non-standard Lax
representation for two special classes of values of the
deformation parameters. In general, we argue that a non-standard
description will involve a pseudo-differential operator of
infinite order.

\newpage

\section{Introduction:}

The search for exactly solvable equations has acquired enormous
importance since the nonlinear Kortweg-de Vries equation was shown
to be integrable \cite{faddeev,gardner}. Exactly solvable
equations, linear or nonlinear, constitute a very special class of
 dynamical systems with many interesting properties. The significance
  of determining new integrable systems, as well as studying their
  properties and solutions, can not be overestimated.
The structure of integrable systems (or partial differential
 equations) is highly restrictive and, in general does not allow
 for deformations. In particular, the presence of arbitrary constant parameters
 (which can not be transformed away by some symmetry) is quite rare. One notable
 exception is the two boson equation \cite{kupershmidt}, which has an arbitrary
 constant parameter present, is known to be integrable. Different values of this
 parameter reduce the
 model to other known soluble models. In this paper, we propose a
 new system of equations, with two arbitrary constant parameters, that is exactly
 soluble and reduces to various other known physical models in different
 limits of these parameters.

Let us recall that the Harry Dym (HD) equation
\cite{kruskal,hereman}
\begin{equation}
u_{t} = (u_{xx}^{-1/2})_{x}\label{hd}\;,
\end{equation}
and the Hunter-Zheng (HZ) equation \cite{hunter}
\begin{equation}
\left(u_{t}+ uu_{x}\right)_{x} =  \frac{1}{2} u_{x}^{2}\label{hz}\;,
\end{equation}
are known to be integrable and, in fact, belong to the same
hierarchy corresponding to negative and positive order flows,
respectively \cite{brunelli}. The HD equation has interest in the study of the
Saffman-Taylor problem which describes the motion of a
two-dimensional interface between a viscous and a non-viscous
fluid \cite{kadanoff}. The HZ equation, on the other hand, arises
in the study of massive nematic liquid crystals and in the study
of shallow water waves \cite{hunter}. In this paper we will show
that the following system of two equations,
\begin{equation}
u_t= \sqrt{2}\left({1-\lambda u_{xx}}\over
\sqrt{2u_{xx}-\alpha-\lambda u_{xx}^2}\right)_x\label{dhd}
\end{equation}
and
\begin{equation}
u_{xxt}=\alpha u_x-2 u_x u_{xx}-u u_{xxx} +{\lambda\over
6}\left(u_x^3\right)_{xx}\label{dhz}\;,
\end{equation}
are integrable for arbitrary values of the constant parameters
$\alpha,\lambda$. Furthermore, they belong to the same hierarchy
corresponding to the negative and positive flows respectively,
much like the HD and the HZ equations. It is interesting to note
from (\ref{dhz}) that when $\alpha\neq 0$, it can be scaled to
unity through the scaling $x\rightarrow {x}/{\alpha}$,
$u\rightarrow {u}/{\alpha}$. Therefore, looking at (\ref{dhz})
alone, it would appear that there are only two meaningful values
for $\alpha$, namely, $\alpha=0,1$. Similarly, under a scaling
$t\rightarrow {t}/{\alpha^{3/2}}$, $x\rightarrow
{x}/{\alpha}$, $u\rightarrow {u}/{\alpha}$, the parameter
$\alpha$ in the dHD equation (\ref{dhd}) can be scaled to unity
when it is not zero. However, there is no scaling which will scale
$\alpha$ to unity (when it is nonzero) simultaneously in both the
equations and, consequently, if the two systems belong to the same
hierarchy, $\alpha$ has to be thought of as an arbitrary
parameter.

We note that, for $\alpha=\lambda=0$, eqs. (\ref{dhd}) and
(\ref{dhz}) reduce respectively to the HD equation (\ref{hd}) and
the HZ equation (\ref{hz}). For lack of a better name, we will
refer to (\ref{dhd}) as the deformed Harry Dym equation (dHD) and
(\ref{dhz}) as the deformed Hunter-Zheng equation (dHZ). (The name
\lq\lq generalized Harry Dym" equation has already been used
earlier in the context of a multi-component Harry Dym equation
\cite{popowicz}.) For $\alpha\not= 0$ and $\lambda=0$ equation
(\ref{dhz}) was studied by Alber et al \cite{alber} and its
solutions contain solitons of the type known as umbilic solitons.
Equation (\ref{dhz}) with $\alpha=1$ and $\lambda\not=0$ has
appeared more recently in the literature \cite{manna} as
describing short capillary-gravity waves and its solutions are
known to become multi-valued in a finite amount of time. While a
matrix Lax pair for this system was provided in \cite{manna} and
solubility of the model was argued based on a map to the
sine-Gordon equation, interesting properties, such as the
bi-Hamiltonian structure, the infinite number of conserved charges
were not discussed at all. Our results clarify these aspects and
provide direct support for the integrability of this new system.
The deformed equation (\ref{dhd}) for arbitrary $\alpha$ and
$\lambda\not=0$ is truly a new integrable system (which to the
best of our knowledge has not been studied in the literature) and
leads to the above mentioned models in different limits. This can,
therefore, be considered as a very rich system, much like the two
boson hierarchy.

Our paper is organized as follows. In section 2, we obtain the
first Hamiltonian structure of the dHD and dHZ equations. In
section 3, a second Hamiltonian structure, compatible with the
first one, is obtained. We show that both the dHD and dHZ
equations are bi-Hamiltonian and, therefore, integrable and belong
to the same hierarchy corresponding to negative and positive
flows. In section 4, we present a Lax pair description of both the
system of equations and construct the conserved charges of
negative order from the Lax operator. We obtain a non-standard Lax
description for the dHD equation for the special values of the
deformation parameters $\lambda = 0, \lambda = -{3}/{\alpha}$.
We argue that, for general values of the parameters, a
non-standard Lax description will involve a pseudo-differential
operator of infinite order. In section 5 we present our
conclusions.

\section{dHZ and dHD as Hamiltonian Systems:}

To describe the dHD and the dHZ equations in a compact manner, let
us introduce the following notation. Let us define
\begin{equation}
F^{2}_{\scriptscriptstyle(\alpha,\lambda)} \equiv 2u_{xx}-\alpha-\lambda u_{xx}^2\;,
\quad A \equiv \frac{1-\lambda u_{xx}}{F}\;.\label{F}
\end{equation}
Then, it follows from the definitions in (\ref{F}) that
\begin{eqnarray}
\kappa \newequiv (1-\alpha\lambda) = \lambda
F^2+(1-\lambda u_{xx})^2 = (\lambda + A^{2}) F^{2}\;,\nonumber\\\noalign{\vskip 5pt}
\frac{F_{x}}{F^{3}} \equal -\frac{1}{\kappa}\,A A_{x}\;,\nonumber\\\noalign{\vskip 5pt}
{u_{xxx}\over F^3}\equal- \frac{1}{\kappa} A_{x}\;.\label{identities}
\end{eqnarray}
These and other relations following from these prove very useful
in the analysis of the two systems.

Given the dHZ equation (\ref{dhz}), we obtain from the definition
in (\ref{F}) that
\begin{equation}
F_t=-\left[F\left(u-{\lambda\over2}u_x^2\right)\right]_x\;.\label{fdhz}
\end{equation}
Similarly, in terms of the new variables in (\ref{F}), the dHD
equation (\ref{dhd}) can be written as
\begin{equation}
u_{t} = \sqrt{2}\, A_{x}\;,\label{dhd1}
\end{equation}
and it follows that, under the evolution of dHD,
\begin{eqnarray}
F_{t} \equal \sqrt{2}\, A\,A_{xxx}\;,\nonumber\\\noalign{\vskip 5pt}
A_{t} \equal -\sqrt{2}\,\kappa\,\frac{A_{xxx}}{F^{3}}\;.\label{fdhd}
\end{eqnarray}
It is now straightforward to note from eqs. (\ref{fdhz}) and
(\ref{fdhd}) that
\begin{equation}
H_{-1}=\sqrt{2}\int dx\, F\label{h-1}
\end{equation}
is conserved under both the dHD and dHZ flows.

The dHZ equation can be obtained from a variational principle,
$\delta\int dtdx\,{\cal L}$, with the Lagrangian density
\begin{equation}
{\cal L}={1\over 2}u_x u_t + {\alpha\over2} u^2 + {1\over 2}uu_x^2
- {\lambda\over 24}u_x^4\;.\label{lagrangian}
\end{equation}
This is a first order Lagrangian density and, consequently, the
Hamiltonian structure can be readily read out, or we can use, for
example, Dirac's theory of constraints \cite{dirac} to obtain the
Hamiltonian and the Hamiltonian operator associated with
(\ref{lagrangian}). The Lagrangian is degenerate and the primary
constraint is obtained to be
\begin{equation}
\Phi=\pi-{1\over 2}u_x\;,\label{primary}
\end{equation}
where $\pi={{\partial{\cal L}}/{\partial u_t}}$ is the canonical
momentum. The total Hamiltonian can be written as
\begin{eqnarray}
H_T \equal\int dx\left(\pi u_t-{\cal L}+\beta\Phi\right)\nonumber\\
\equal\int dx\left[- {\alpha\over2}u^2 - {1\over 2}uu_x^2 +
{\lambda\over 24}u_x^4+\beta\left(\pi-{1\over
2}u_x\right)\right]\;,\label{ht}
\end{eqnarray}
where $\beta$ is a Lagrange multiplier field. Using the canonical
Poisson bracket relation
\begin{equation}
\{u(x),\pi(y)\}=\delta(x-y)\;,\label{poisson}
\end{equation}
with all others vanishing, it follows that the requirement of the
primary constraint to be stationary under time evolution,
\[
\{\Phi(x),H_T\}=0\;,
\]
determines the Lagrange multiplier field $\beta$ in (\ref{ht}) and
the system has no further constraints.

Using the canonical Poisson bracket relations (\ref{poisson}), we
can now calculate
\begin{equation}
K(x,y)\equiv\{\Phi(x),\Phi(y)\}={1\over
2}\partial_y\delta(x-y)-{1\over
2}\partial_x\delta(x-y)\;.\label{kpoisson}
\end{equation}
This shows that the constraint (\ref{primary}) is second class and
that the Dirac bracket between the basic variables has the form
\[
\{u(x),u(y)\}_D=\{u(x),u(y)\}-\int
dz\,dz'\{u(x),\Phi(z)\}J(z,z')\{\Phi(z'),u(y)\}=J(x,y)\;,\
\]
where $J$ is the inverse of the Poisson bracket of the constraint
(\ref{kpoisson}),
\[
\int dz\,K(x,z) J(z,y)=\delta(x-y)\,.
\]
This last relation determines
\[
\partial_x J(x,y)=\delta(x-y)\;,
\]
or
\[
J(x,y)={\cal D}_1\delta(x-y)\;,
\]
where
\begin{equation}
{\cal D}_1=\partial^{-1}\;,\label{d1}
\end{equation}
and can be thought of as the alternating step function in the
coordinate space. We can now set the constraint (\ref{primary})
strongly to zero in (\ref{ht}) to obtain
\begin{equation}
H_2\equiv -H_T =\int dx\left({\alpha\over2}u^2+{1\over
2}uu_x^2-{\lambda\over 24}u_x^4\right)\;.\label{h2}
\end{equation}
Therefore, the dHZ equation can be written in the Hamiltonian form
\[
u_t={\cal D}_1{\delta H_2\over\delta u}\;,
\]
with ${\cal D}_1$  and $H_{2}$ given by (\ref{d1}) and (\ref{h2}),
respectively.

From the results in \cite{brunelli} we know that the HD and HZ
equations belong to the same hierarchy of equations. Here, too, we
will see that both the dHD and the dHZ equations belong to the
same hierarchy. In particular, we note that
\[
u_t={\cal D}_1{\delta H_{-1}\over\delta u}\;,
\]
with $H_{-1}$, given by (\ref{h-1}), yields the deformed Harry Dym
equation (\ref{dhd1}). As a result, the dHD equation also is
Hamiltonian with the same Hamiltonian structure of the dHZ
equation in (\ref{d1}) and, consequently, has a Lagrangian
description given by
\[
 {\cal L} = \frac{1}{2} u_{x}u_{t} - \sqrt{2}\, F\;.
\]
The reader can easily check that $H_2$ is also conserved by both
the dHD and dHZ equations, much like $H_{-1}$. Note that for
$\alpha=\lambda=0$, these two charges reduce to the corresponding
ones in the HD and HZ systems.

\section{dHD and dHZ as bi-Hamiltonian Systems:}

It is well known that a system can be shown to be integrable if it
is bi-Hamiltonian \cite{magri,olver}. This corresponds to the
system having a Hamiltonian description with two distinct
Hamiltonian structures that are compatible. Therefore, we try to
see if the dHD and the dHZ equations can be described as
bi-Hamiltonian systems. For this, we have to find a second
Hamiltonian description for the two systems.

In order to determine a second Hamiltonian structure for the two
systems, we recall \cite{brunelli} that, when $\alpha=\lambda=0$,
the corresponding $H_{-1}$ is a Casimir of the second Hamiltonian
structure for the HD and HZ systems, namely,
\begin{equation}
{\cal D}_2^{\scriptscriptstyle(\alpha=\lambda=0)}=\partial^{-2}u_{xx}\,\partial^{-1}+
\partial^{-1}u_{xx}\,\partial^{-2}\label{d2hd}
\end{equation}
satisfies
\[
{\cal D}_{2}^{\scriptscriptstyle(\alpha=\lambda=0)}\, \frac{\delta H_{-1}^{\scriptscriptstyle(\alpha=\lambda=0)\!\!\!\!\!\!\!\!\!\!}}{\delta u} = 0\;.
\]
Since we know $H_{-1}$ for the deformed systems which is a
generalization of  $H_{-1}^{\scriptscriptstyle(\alpha=\lambda=0)}$, we look for a
Hamiltonian structure for which it is a Casimir. With some work,
it can be determined that $H_{-1}$ given in (\ref{h-1}) is a
Casimir of
\begin{equation}
{\cal D}_2\equiv {\cal D}_{2}^{\scriptscriptstyle(\alpha,\lambda)} ={1\over
2}\left(\partial^{-2}F^2\,\partial^{-1}+\partial^{-1}F^2\,\partial^{-2}\right)+
\lambda\,\partial^{-2}u_{xxx}\,\partial^{-1}u_{xxx}\partial^{-2}
\;.\label{d2}
\end{equation}
Note that this structure reduces to (\ref{d2hd}) when
$\alpha=\lambda=0$. The skew symmetry of this Hamiltonian
structure is manifest. The proof of the Jacobi identity for this
structure as well as its compatibility with ${\cal D}_{1}$ in
(\ref{d1}) can be determined through the standard method of
prolongation described in ref. \cite{olver}, which we discuss
briefly.

Performing the change of variables
\[
w=u_{xx}\;,\nonumber
\]
the Hamiltonian structures (\ref{d1}) and (\ref{d2}) assume the
forms
\begin{eqnarray}
{\cal D}_1 \equal\partial^3\;,\nonumber\\\noalign{\vskip 5pt} {\cal
D}_2 \equal{1\over2}\left(F^2\partial+\partial F^2\right)+\lambda
w_x\,\partial^{-1}w_x\;.\nonumber
\end{eqnarray}
We can construct the two bivectors associated with the two
structures as
\begin{eqnarray}
\Theta_{{\cal D}_1}\equal{1\over 2}\int dx\,\left\{\theta\wedge{\cal
D}_1\theta\right\}={1\over2}\int
dx\,\theta\wedge\theta_{xxx}\;,\nonumber\\\noalign{\vskip 5pt}
\Theta_{{\cal D}_2}\equal{1\over 2}\int dx\,\left\{\theta\wedge{\cal
D}_2\theta\right\}\nonumber\\
\equal{1\over2}\int
dx\,\left\{-\alpha\,\theta\wedge\theta_x+2w\,\theta\wedge\theta_x-\lambda\,
w^2\theta\wedge\theta_x+\lambda\, w_x\,\theta\wedge(\partial^{-1}
w_x\,\theta)\right\}\;.\nonumber
\end{eqnarray}
Using the prolongation relations,
\begin{eqnarray}
\hbox{\bf pr}\,{\vec v}_{{\cal D}_1{\theta}} (w)
\equal\theta_{xxx}\nonumber\;,\\\noalign{\vskip 5pt} \hbox{\bf
pr}\,{\vec v}_{{\cal D}_1{\theta}} (w^2) 
\equal 2w\,\hbox{\bf
pr}\,{\vec v}_{{\cal D}_1{\theta}}
(w)\nonumber\;,\\\noalign{\vskip 5pt} \hbox{\bf pr}\,{\vec
v}_{{\cal D}_1{\theta}} (w_x) \equal\left(\hbox{\bf pr}\,{\vec
v}_{{\cal D}_1{\theta}} (w)\right)_x\;,\nonumber\\\noalign{\vskip
7pt} \hbox{\bf pr}\,{\vec v}_{{\cal D}_2{\theta}} (w)
\equal-\alpha\,\theta_x+2w\,\theta_x-\lambda\,
w^2\theta_x+w_x\,\theta-\lambda\,ww_x\,\theta+\lambda
w_x\,(\partial^{-1} w_x\,\theta)\nonumber\;,\\\noalign{\vskip 5pt}
\hbox{\bf pr}\,{\vec v}_{{\cal D}_2{\theta}} (w^2)
\equal2w\,\hbox{\bf pr}\,{\vec v}_{{\cal D}_2{\theta}}
(w)\nonumber\;,\\\noalign{\vskip 5pt} \hbox{\bf pr}\,{\vec
v}_{{\cal D}_2{\theta}} (w_x) \equal\left(\hbox{\bf pr}\,{\vec
v}_{{\cal D}_2{\theta}} (w)\right)_x\;,\label{prolongation}
\end{eqnarray}
it is straightforward to show that the prolongation of the
bivector $\Theta_{{\cal D}_2}$ vanishes,
\[
\hbox{\bf pr}\,{\vec v}_{{\cal D}_2\theta}\left(\Theta_{{\cal
D}_2}\right)=0\;,
\]
implying that ${\cal D}_2$ satisfies  Jacobi identity.  Using
(\ref{prolongation}), it also follows that
\[
\hbox{\bf pr}\,{\vec v}_{{\cal D}_1\theta}\left(\Theta_{{\cal
D}_2}\right)+\hbox{\bf pr}\,{\vec v}_{{\cal
D}_2\theta}\left(\Theta_{{\cal D}_1}\right)=0\;,
\]
showing that ${\cal D}_1$ and ${\cal D}_2$ are compatible. Namely,
not only are ${\cal D}_{1}, {\cal D}_{2}$ genuine Hamiltonian
structures, any arbitrary linear combination of them is as well.
Any physical system that is Hamiltonian with respect to these two
structures, therefore, defines a pencil system and is integrable.
It is worth pointing out here that when $\alpha=\lambda=0$, the
second Hamiltonian structure (\ref{d2hd}) represents the
centerless Virasoro algebra \cite{das} (with dimension zero
operators). The structure in (\ref{d2}) appears to be a highly
nonlocal generalization of this algebra, but we are not familiar
with any study of such an algebra in the literature.

To show that the dHD and dHZ are bi-Hamiltonian systems, we note
that the charges
\begin{eqnarray}
H_1 \equal \int dx\, u_x^2\label{h1}\;,\\\noalign{\vskip 5pt}
H_{-2} \equal \frac{1}{2\kappa} \int dx\,F A_{x}^{2}\label{h-2}
\end{eqnarray}
are also conserved by both the dHZ and dHD equations. While the
charge in (\ref{h1}) is the unmodified charge of the HD and HZ
systems (this seems to be a simple coincidence), the charge in
(\ref{h-2}) is a true generalization of the corresponding charge
of the HD and HZ systems. With these charges, it is easy to see
that the dHZ equation can be written in a truly bi-Hamiltonian
form
\[
u_t={\cal D}_1{\delta H_2\over\delta u}={\cal D}_2{\delta
H_1\over\delta u}\;.
\]
Similarly, the dHD equation can also be written in the
bi-Hamiltonian form
\[
u_t={\cal D}_1{\delta H_{-1}\over\delta u}={\cal D}_2{\delta
H_{-2}\over\delta u}\;.
\]
Thus, we see that both the dHD as well as dHZ systems are
bi-Hamiltonian with the same two compatible Hamiltonian structures
and are, therefore, integrable.

\section{The Lax Representation:}

When a system is  bi-Hamiltonian, we can naturally define a
hierarchy of commuting flows through the relation
\begin{equation}
u_t=K_n[u]={\cal D}_1{\delta H_{n+1}\over\delta u}={\cal
D}_2{\delta H_{n}\over\delta u}\;,\quad
n=0,1,2,\dots\;.\label{eqrecursion}
\end{equation}
In the present case, both the Hamiltonian structures have
Casimirs. We have already seen that $H_{-1}$ is a Casimir of
${\cal D}_{2}$ and it can be checked that the trivial Hamiltonian
\begin{equation}
H_{0} = \int dx\,u_{xx}\label{h0}
\end{equation}
formally defines the Casimir of ${\cal D}_{1}$ (namely, if we
write formally $\frac{\delta H_{0}}{\delta u} = \partial^{2}$, it
is annihilated by ${\cal D}_{1}$). As a result, the system of
flows can be extended to both positive and negative integer values
for $n$. In this way, we see that much like in the HD and HZ
systems \cite{brunelli}, the dHD and dHZ systems also belong to the same hierarchy
corresponding to the negative and positive flows.

Let us introduce the recursion operator following from the two
Hamiltonian structures as
\[
R={\cal D}_2{\cal D}_1^{-1}\;.
\]
Then, it follows from (\ref{eqrecursion}) that
\[
K_{n+1}=R\,K_n\;,
\]
and
\begin{equation}
{\delta H_{n+1}\over\delta u}=R^\dagger{\delta H_{n}\over\delta
u}\;,\label{hrecursion}
\end{equation}
where
\begin{equation}
R^\dagger=\partial^{-1}u_{xx}\,\partial^{-1}+
(-\alpha+u_{xx})\,\partial^{-2}-\lambda\,
u_{xx}\,\partial^{-1}u_{xx}\,\partial^{-1}\label{rdagger}
\end{equation}
is the adjoint of $R$. The conserved charges for the hierarchy
can, of course, be determined in principle recursively from
(\ref{hrecursion}). However, in practice, integrating the
recursion relation is highly nontrivial. Therefore, we look for a
Lax representation for the system of dHD and dHZ equations which
will allow us to construct the conserved charges directly.

It is well known \cite{okubo,recursion} that for a bi-Hamiltonian
system of evolution equations, $u_t=K_n[u]$, a natural Lax
description
\[
{\partial M\over\partial t}=[M,B]\;,
\]
is easily obtained where, we can identify
\begin{eqnarray}
M \newequiv R^\dagger\;,\nonumber\\
\noalign{\vskip 5pt} B
\newequiv K'_n\;.\nonumber
\end{eqnarray}
Here $K'_n$ represents the Fr\'echet derivative of $K_n$, defined
by
\[
K'_n[u]\,v=\frac{d\ }{d\epsilon}\,K_n[u+\epsilon
v]\Big|_{\epsilon=0}\Big.\;.
\]
For the dHD and dHZ system of equations in (\ref{dhd}) and
(\ref{dhz}) respectively, we have
\begin{eqnarray}
B^{\hbox{\scriptsize\,dHD}}\newequiv
K_{-2}=\sqrt{2}\,\kappa\,\partial^2 F^{-3}\partial
\;,\nonumber\\\noalign{\vskip 5pt}
B^{\hbox{\scriptsize\,dHZ}}\newequiv
K_1=(-\alpha+u_{xx})\,\partial^{-1}+\partial\left(u-{\lambda\over
2}u_x^2\right)
\;.\nonumber
\end{eqnarray}
The two systems have the same $M=R^{\dagger}$ given in
(\ref{rdagger}). It can now be checked that
\begin{eqnarray}
\frac{\partial M}{\partial t} \equal \left[M, B^{\hbox{\scriptsize
\,dHD}}\right]\;,\nonumber\\\noalign{\vskip 5pt}
\frac{\partial M}{\partial t} \equal \left[M, B^{\hbox{\scriptsize
\,dHZ}}\right]\;,\label{lax}
\end{eqnarray}
do indeed generate the dHD and the dHZ equations and, thereby,
provide a Lax pair for the system.

One of the advantages of a Lax representation is that they
directly give the conserved charges of the system. From the
structure of (\ref{lax}), it follows that ${\rm
Tr}\,M^{\frac{2n+1}{2}}$ are conserved, where ``Tr" represents
Adler's trace \cite{adler}. We note that
\begin{eqnarray}
\hbox{Tr}M^{2n+1\over 2} \equal 0\;,\quad
n\ge1\nonumber\\\noalign{\vskip 5pt} \hbox{Tr}M^{1\over 2} \equal
\int dx\, F\;,\nonumber\\\noalign{\vskip 5pt}
\hbox{Tr}M^{-{1\over2}} \equal -{1\over 2\kappa}\int dx\,F
A_{x}^{2}\;,\nonumber\\\noalign{\vskip 5pt}
\hbox{Tr}M^{-{3\over2}} \equal \frac{3}{\kappa} \int dx\,\left(4\,
\frac{A_{xx}^{2}}{F} + \frac{1}{\kappa} F
A_{x}^{4}\right)\;,\nonumber\\
&\vdots&\;.\label{charges}
\end{eqnarray}
The first two nontrivial charges correspond respectively to
$H_{-1},H_{-2}$ given in eqs. (\ref{h-1}) and (\ref{h-2}),
constructed earlier by brute force. In fact, all $H_{-n-1}$ with
positive $n\geq 0$ can be constructed from ${\rm
Tr}\,M^{-\frac{2n-1}{2}}$ and by construction (namely, because of
the nature of (\ref{lax})), they are conserved under both the dHD
and dHZ flows. Unfortunately, as is clear from (\ref{charges}),
this procedure does not yield the charges $H_{n}$ with positive
integer values. These are, in general, non-local and even in the
HD and HZ case, construction of these charges relies primarily on
the recursion relation (\ref{hrecursion}). It remains an
interesting question to construct these charges in a more direct
manner.

The Harry Dym equation has a Gelfand-Dikii representation for the
Lax pair, while the HZ equation does not. We will now discuss the
existence of such a Lax representation for the dHD equation. A
spectral problem associated with the dHD equations can be obtained
from the recursion relation (\ref{hrecursion}) (see \cite{camassa}
and references therein) as follows. Introducing a spectral
parameter $\mu$ and defining
\[
\psi^2(x,t,\mu)=\sum_{n=0}^\infty \mu^{n}{\delta H_{n}\over\delta
u}\;,
\]
we note that the recursion relation (\ref{hrecursion}) can be
written compactly as (recall that $H_{0}$ is a Casimir of ${\cal
D}_{1}$)
\[
({\cal D}_1- \mu {\cal D}_2)\psi^2=0\;,
\]
or
\begin{equation}
(1 - \mu R^\dagger)\,\psi^2= 0\;,\label{eigenproblem}
\end{equation}
which defines an eigenvalue problem for the eigenfunction $\psi^2$
with eigenvalue ${1}/{\mu}$. A linear eigenvalue problem can
be derived from this if we can factorize the operator $(1 - \mu
R^{\dagger})$.

Let us note that the operator $R^{\dagger}$ in (\ref{rdagger}) can
be rewritten in the form
\begin{equation}
R^{\dagger} = \frac{1}{2}\left[\partial^{-1}
\Bigl(F^{2}_{\scriptscriptstyle(\alpha,\lambda)} + X\Bigr)\partial^{-1} + \Bigl(F^{2}_{\scriptscriptstyle(\alpha,\lambda)}
+ X\Bigr)\partial^{-2}\right]\;,\label{rnew}
\end{equation}
where
\[
X = \sum_{n=1} X_{n} \partial^{-n}\;,
\]
and the coefficients $X_{n}$ can be determined recursively to be
\begin{eqnarray}
X_{1} \equal 0\;,\nonumber\\\noalign{\vskip 5pt}
X_{2} \equal \frac{\lambda}{4} u_{xxx}^{2}\;,\nonumber\\\noalign{\vskip 5pt}
X_{3} \equal - \frac{1}{2} X_{2,x}\;,\nonumber\\
&\vdots &\;. \label{X}\nonumber
\end{eqnarray}
When $\lambda = 0$, it follows that $X=0$ and that
$F^{2}_{\scriptscriptstyle(\alpha,\lambda=0)}$ is a simple function. In this case, the
eigenvalue problem (\ref{eigenproblem}) can be factorized as
\[
(1-R^{\dagger})\,\psi^{2} = 2\,\partial^{-1}\phi^{-2}\,\partial\,\phi^{3}
\left(\partial^{2} - \frac{\mu}{4}\, F^{2}_{\scriptscriptstyle(\alpha,\lambda=0)}\right)\phi = 0\;,
\]
where we have identified
\[
\phi^{2} = (\partial^{-2}\psi^{2})\;.
\]
This shows that if the linear equation
\[
\left(\partial^{2} - \frac{\mu}{4} F^{2}_{\scriptscriptstyle(\alpha,\lambda=0)}\right)\phi = 0
\]
is satisfied, then (\ref{eigenproblem}) will hold and this
identifies the Lax operator for the system to be
\[
L = \frac{1}{F^{2}_{\scriptscriptstyle(\alpha,\lambda=0)}}\, \partial^{2}\;.
\]
In fact, it can be readily checked that when $\lambda = 0$, the
hierarchy of dHD equations can be obtained from the non-standard
Lax equation
\[
\frac{\partial L}{\partial t_{n}} = {4\sqrt{2}}\left[L, (L^{(2n-1)/2})_{\geq
2}\right]\;.
\]
The conserved quantities for this system can be obtained from
${\rm Tr}\,L^{(2n-1)/2}$, $n=0,1,2,\dots$.

On the other hand, when $\lambda\neq 0$, the coefficients $X_{n}$
are nontrivial and $X$ represents a pseudo-differential operator.
The factorization, in such a case, is not so simple and, in
principle would involve an infinite series of terms. For arbitrary
values of $\kappa$, the terms in the series can possibly be
determined recursively. However, this is not very interesting. We
simply note here that for the special value $\kappa = 4$, the
infinite series of terms seems to have a simpler compact form and
the Lax operator, in such a case, has the form
\[
L_{\kappa=4} = \frac{1}{F}\partial \frac{1}{F} \partial -
\frac{1}{4} A_{x}\partial^{-1}A_{x}\partial\;,
\]
and, in this case, the hierarchy of dHD equations can be obtained
from the non-standard Lax representation
\[
\frac{\partial L_{\kappa=4}}{\partial t_{n}} = {4\sqrt{2}}\left[L_{\kappa=4},
\left(L_{\kappa=4}^{(2n-1)/2}\right)_{\geq 2}\right]\;.
\]
The conserved quantities, in this case, also follow from ${\rm
Tr}\,L_{\kappa=4}^{(2n-1)/2}$ and up to multiplicative factors, they have the
forms given in (\ref{charges}) with $\kappa=4$. A simple Lax
description for arbitrary $\lambda$, however, remains an open
question.

\section{Conclusion:}

In this paper, we have studied the general system of dHD and dHZ
equations. We have shown that  both these systems are
bi-Hamiltonian and, therefore, integrable and belong to the same
hierarchy corresponding to negative and positive flows. The Lax
pair for the two system of equations have been derived and
conserved charges corresponding to negative integer values follow
from the Lax operator. A simple construction of the charges for
positive integer values remains an open question. For $\lambda=0$,
we have constructed a non-standard Lax representation for the dHD
equation, which involves a purely differential Lax operator. For
arbitrary values of $\lambda$, we have argued that a non-standard
Lax representation will necessarily involve a Lax operator which
is a pseudo-differential operator of infinite order. For the
particular case of $\kappa=4$, however, this takes a simpler
compact form.

\section*{Acknowledgments}

One of us (AD) would like to thank the members of the physics
Departments at UFSC (Brazil) and Wroc\l aw University (Poland) for
hospitality, where parts of this work was carried out. This work
was supported in part by US DOE grant no. DE-FG-02-91ER40685 as
well as by NSF-INT-0089589.

\end{document}